%% file: mass-prl.tex
\documentclass[prl,twocolumn,showpacs,preprintnumbers,amsmath,amssymb,floatfix,superscriptaddress]{revtex4}

\usepackage{graphicx}
\usepackage{dcolumn}
\usepackage{bm}

\newcommand{\um}   {\ensuremath{\mathrm{\mu m}}}
\newcommand{\TeV}  {\ensuremath{\mathrm{Te \kern-0.1em V}}}
\newcommand{\GeV}  {\ensuremath{\mathrm{Ge \kern-0.1em V}}}
\newcommand{\GeVc} {\ensuremath{\mathrm{Ge \kern-0.1em V/}c}}
\newcommand{\GeVcc}{\ensuremath{\mathrm{Ge \kern-0.1em V/}c^2}}
\newcommand{\MeV}  {\ensuremath{\mathrm{Me \kern-0.1em V}}}
\newcommand{\MeVc} {\ensuremath{\mathrm{Me \kern-0.1em V/}c}}
\newcommand{\MeVcc}{\ensuremath{\mathrm{Me \kern-0.1em V/}c^2}}

\begin{document}


\title{\boldmath
       Measurement of $b$ hadron masses in 
       exclusive $J/\psi$ decays with the CDF detector
       \unboldmath}

\input{author_list_cdf_feb_2005}

\collaboration{ CDF Collaboration }
\noaffiliation

\date{\today}
\begin{abstract}
\noindent
We measure the masses of $b$ hadrons in exclusively reconstructed final states 
containing a $J/\psi\to\mu^-\mu^+$ decay using $\rm 220~pb^{-1}$ of
data collected by  the CDF II experiment. 
We find:
\begin{center}
\vspace{-0.5cm}
\[
\begin{array}{lclcrcrcr}
\multicolumn{3}{l}{m(B^+)}           &=&5279.10&\pm&0.41\,_{(stat.)}\;&\pm& 0.36\,_{(sys.)}~\MeVcc,\\
\multicolumn{3}{l}{m(B^0)\,}         &=&5279.63&\pm&0.53\,_{(stat.)}\;&\pm& 0.33\,_{(sys.)}~\MeVcc,\\
\multicolumn{3}{l}{m(B^0_s)\,}       &=&5366.01&\pm&0.73\,_{(stat.)}\;&\pm& 0.33\,_{(sys.)}~\MeVcc,\\
\multicolumn{3}{l}{m(\Lambda^0_b)\,} &=&5619.7 &\pm&1.2 \,_{(stat.)}\;&\pm& 1.2 \,_{(sys.)}~\MeVcc.\\
m(B^+)&-&m(B^0)    &=&  -0.53&\pm&0.67\,_{(stat.)}  &\pm& 0.14\,_{(sys.)}~\MeVcc,\\
m(B^0_s)&-&m(B^0)  &=&  86.38&\pm&0.90\,_{(stat.)}  &\pm& 0.06\,_{(sys.)}~\MeVcc,\\
m(\Lambda^0_b)&-&m(B^0)&=&339.2&\pm& 1.4\,_{(stat.)} &\pm&0.1\,_{(sys.)}~\MeVcc.
\end{array}
\]
\end{center}
%
The measurements of the $B^0_s$, $\Lambda^0_b$ mass, 
$m(B^0_s)-m(B^0)$ and $m(\Lambda^0_b)-m(B^0)$ mass difference 
are of better precision than the current world averages. 
\end{abstract}

\maketitle

In the Standard Model, Quantum Chromodynamics (QCD) is the theory
describing strong interactions between objects of color charge~\cite{QCD}.
Hadrons are colorless particles made up of strongly-interacting
constituents.
Hadron masses are fundamental physical observables and 
their study is the spectroscopy of quark systems bound by QCD. 
The heaviest known hadrons contain a bottom or $b$
quark~\cite{BQUARK}. Lattice QCD calculations 
predict the hadron mass spectra from first principles. Comparing 
these calculations with experimental data is an essential test of QCD.  
Recent advances in lattice QCD, using unquenched methods, will soon allow 
mass predictions with close to the experimental accuracy~\cite{LQCD2}
and we look forward to these new calculations to compare with our results.
The most precise predictions are those of $b$ hadron mass differences.
We present here the most precise individual measurements, to date, 
for the masses of  $B^+$, $B^0$, $B^0_s$ and $\Lambda^0_b$.

The data used in this analysis were obtained with the Collider
Detector at Fermilab (CDF II) operating at the
$\sqrt{s} = \rm 1.96~\TeV$ Tevatron $p\bar{p}$-collider.
The data were collected between February 2002 and August 2003 
and corresponds to an integrated luminosity of $\rm 220~pb^{-1}$. 
The CDF II detector is described in detail elsewhere~\cite{CDFDESC}. 
This analysis relies on the tracking
system and the muon detectors.  The tracking system is comprised of a silicon
micro strip vertex detector (SVX II)~\cite{SVXII} and a drift chamber 
operating in a $\rm 1.4~T$ solenoidal magnetic field.
The SVX II system consists of 5 concentric 
silicon layers made of double-sided silicon covering the radii from 
$\rm 2.5~cm$ to $\rm 10.6~cm$.  
The impact parameter resolution is about $40~\mu\rm m$, 
including a $30~\mu\rm m$ contribution from the beamspot.
The Central Outer Tracker (COT)~\cite{COT} is an open cell drift
chamber measuring $\rm 310~cm$ in length, with an inner radius of 
$\rm 41~cm$ extending to a radius of $\rm 138~cm$ and provides
a large lever arm for curvature measurements. 
Each cell contains a plane of 12 sense wires tilted
by $35^\circ$ with respect to the radial direction to compensate for the 
drift Lorentz-angle. 
The COT is segmented radially into eight superlayers. 
For superlayers 1, 3, 5 and 7 wires form a $\pm 2^\circ$ stereo angle with
respect to the beam direction, while for superlayers 2, 4, 6 and 8 
wires are oriented along the beam direction.
The measured momentum resolution is $\sigma (p_T)/p_T\sim 0.15\% p_T/(\GeVc)$.
Muon detectors consist of multi-layer drift chambers located around 
the outside of the calorimeters~\cite{CMU}.  
The central muon system covers a range in pseudorapidity of $|\eta|<0.6$. 
The central muon extension extends the pseudorapidity range to 
$0.6<|\eta|<1.0$.

Data are selected with a three-level trigger system.
The Level 1 portion of the dimuon trigger uses the
extremely fast tracker~\cite{XFT}, providing a coarse track
reconstruction based on fast digitization of drift chamber signals.
Only tracks with a measured transverse momentum larger than $1.5~\GeVc$ 
are considered further. Two such tracks, 
matched to distinct hits in the muon systems, 
are required to pass the Level 1 dimuon trigger. 
No additional requirements are made at Level 2.
At Level 3, a detailed reconstruction is performed and opposite 
sign dimuon events with an invariant mass in the range 
$2.7-4.0~\GeVcc$ are accepted and written to tape.
Stored events are reconstructed using the full set of calibrations.


The following $b$ hadron decay modes are reconstructed: $B^+\to
J/\psi K^+$, $B^0 \to J/\psi K^{*0}$, $B^0_s \to J/\psi \phi$, $B^0 \to J/\psi
K^0_S$ and $\Lambda^0_b \to J/\psi \Lambda^0$.  
The daughter particles are reconstructed in the decay modes 
$K^{*0}\to K^+\pi^-$, $\phi \to K^+K^-$, $K^0_S\to\pi^+\pi^-$ 
and $\Lambda^0\to p\pi^-$. 
Charge conjugate modes are included implicitly.
To reconstruct a given final state we try all possible combinations of
particle hypotheses, since hadronic particle identification
capabilities are limited. 
For a given particle hypothesis tracks are corrected for energy loss
with the corresponding mass assigned to the track~\cite{IVAN}.
The correction procedure~\cite{AKORN} makes use of the material
information in a {\sc geant}~\cite{GEANT} description of the CDF detector. 
Material is integrated only at radii larger than
that of the reconstructed decay vertex of long-lived $K^0_S$ and
$\Lambda^0$ particles.   
High track quality is ensured by requiring at least 20
axial and 16 stereo hits in the COT. 
To ensure a precise measurement of the $b$ hadron decay vertex, 
only tracks with at least 3 axial SVX hits are
considered.  The SVX hit requirement is not applied to the
daughter tracks from $K^0_S$ and $\Lambda^0$ which tend to 
decay outside the silicon tracker.
A muon is reconstructed from tracks matched to track stubs in the 
muon chambers.

The mass reconstruction begins by constraining the two triggered,
oppositely charged muons to a common vertex. 
Candidates with a resulting dimuon mass within $80~\MeVcc$ of the 
world average $J/\psi$ mass~\cite{PDG2000} are selected. 
A $p_T$ threshold of $400~\MeVc$ is required on all tracks,
except $\Lambda^0$ daughters, for which all available tracks are used.
A uniform threshold of $2~\GeVc$ is imposed on the momentum 
transverse to the beam direction of $K^+$, $K^{*0}$ and $\phi$ candidates.
Mass windows of $80~\MeVcc$, $10~\MeVcc$ and $40~\MeVcc$ around the 
world average masses~\cite{PDG2000} are required to select 
$K^{*0}$, $\phi$ and $K^0_S$, respectively. 
Combinations with a $p\pi$ mass between $1.10~\GeVcc$ and
$1.13~\GeVcc$ are selected as $\Lambda^0$ candidates.  
The $K^0_S$ and $\Lambda^0$ flight directions are reconstructed as the
vector connecting the $J/\psi$ and the $K^0_S$ or $\Lambda^0$ vertices.
We require the $K^0_S$ and $\Lambda^0$ momentum vectors to be within 
$0.25^\circ$ and $0.57^\circ$ of their flight direction, respectively.
A cut on the $b$ hadron transverse momentum $p_T \ge 6.5~\GeVc$ is
applied. For $b$ hadrons, $c\tau$ is in the range 
$\sim400-500~\mu\rm m$, where $\tau$ is the proper lifetime. 
In order to reduce background, the 2-dimensional decay length of the
$b$ hadron, $L_{xy}$, 
defined as $L_{xy}=\frac{\vec{X}\cdot\vec{p}_T}{|\vec{p}_T|}$, is
required to exceed $100~\um$, where $\vec{X}$ is the vector
between the production vertex and the decay vertex of the $b$ hadron.


\begin{figure}[!h]
\begin{center}
\includegraphics[width=0.45\textwidth]{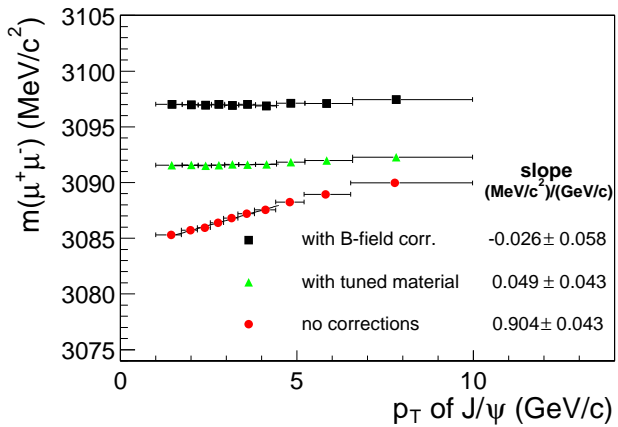}
\end{center}
\caption{The reconstructed mass of dimuons from $J/\psi$ decays, as a
function of transverse momentum. The three lines represent various
stages of corrections: the solid circles indicate no
correction, the triangles add the material tuning and the squares show all
corrections including the magnetic field scale.}
\label{fig:mass-calibration}
\end{figure}
To calibrate the momentum scale of the CDF tracking system, three values
must be determined: the energy lost by a track when passing through 
the material in the inner detector, the radius of the tracker 
(for the track curvature measurement) and the strength of the magnetic
field (for the track curvature-to-momentum conversion).  
The effect of the tracker radius is indistinguishable from the
magnetic field strength in the calibration, so we neglect the tracker
radius and describe the procedure in terms of a magnetic field calibration.  
We use a sample of over 1 million, inclusive $J/\psi\to\mu\mu$ decays
to calibrate the track energy loss and magnetic field.
An underestimate of the material results in undercorrected energy loss
and introduces a dependence of the reconstructed dimuon mass on $p_T$. 
The $p_T$ dependence is a signature for inadequate material
assessment, as an incorrect value of the magnetic field produces a 
shift in invariant mass independent of the $p_T$ of the 
reconstructed particle.
The first calibration step tunes the amount of material to remove
the momentum dependence. 
Next the magnetic field is scaled so that the reconstructed
$J/\psi\to\mu\mu$ mass agrees with the world average. 
The effect of the calibration steps on the momentum dependence of the 
dimuon mass is illustrated in Figure~\ref{fig:mass-calibration}. 
Final state radiation in the decay of the $J/\psi$ leads to an
asymmetry in the otherwise Gaussian distribution of the measured dimuon
invariant mass. We use a Monte Carlo simulation to correct the resulting
bias in bins of $J/\psi$ $p_T$ during calibration.

\begin{figure*}[!t,b]
 \begin{center}
 \begin{minipage}[!h]{\textwidth}
 \begin{minipage}[!h]{0.4\textwidth}
   \includegraphics[width=\textwidth]{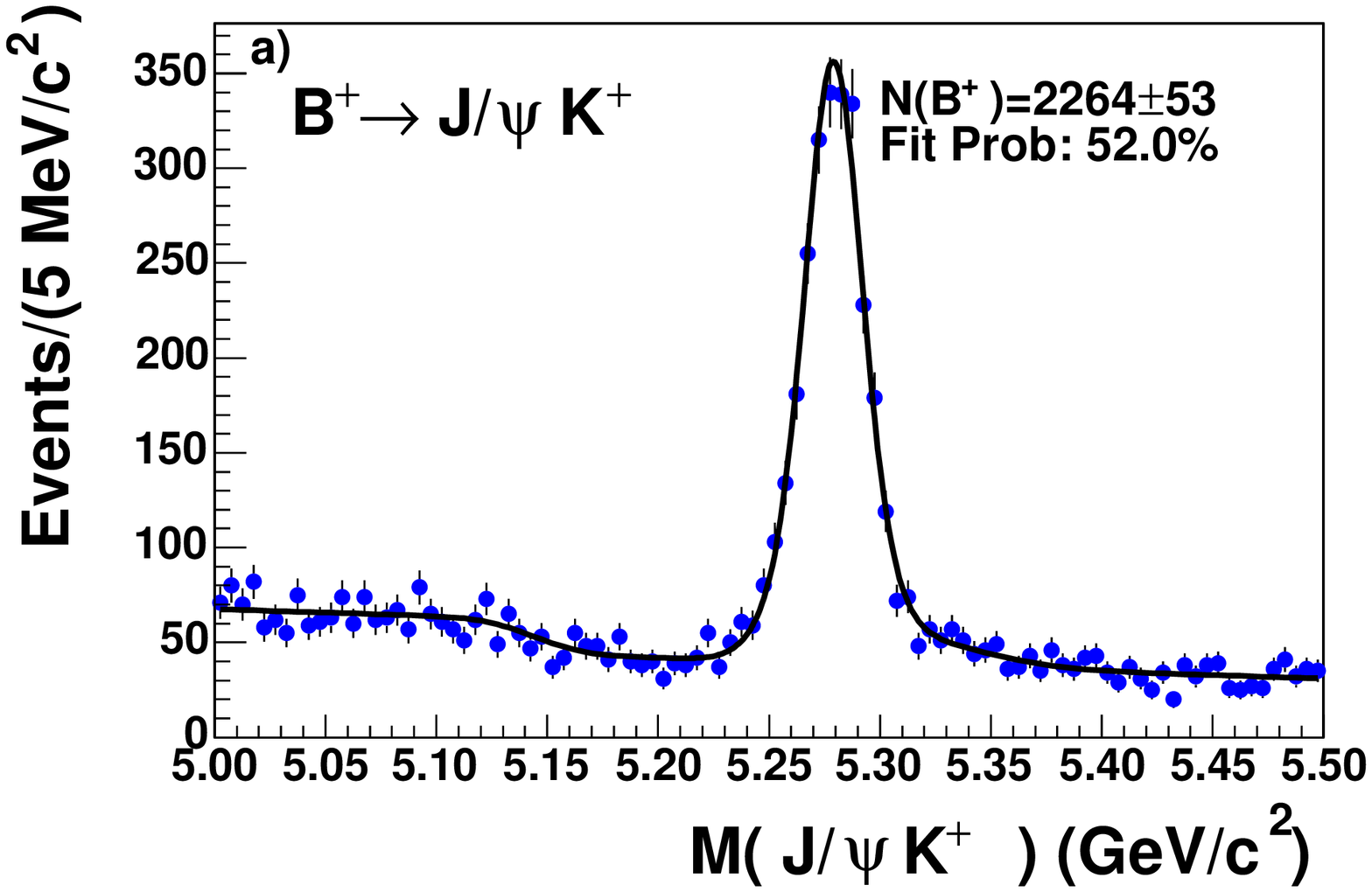}
 \end{minipage}
 \begin{minipage}[!h]{0.4\textwidth}
   \includegraphics[width=\textwidth]{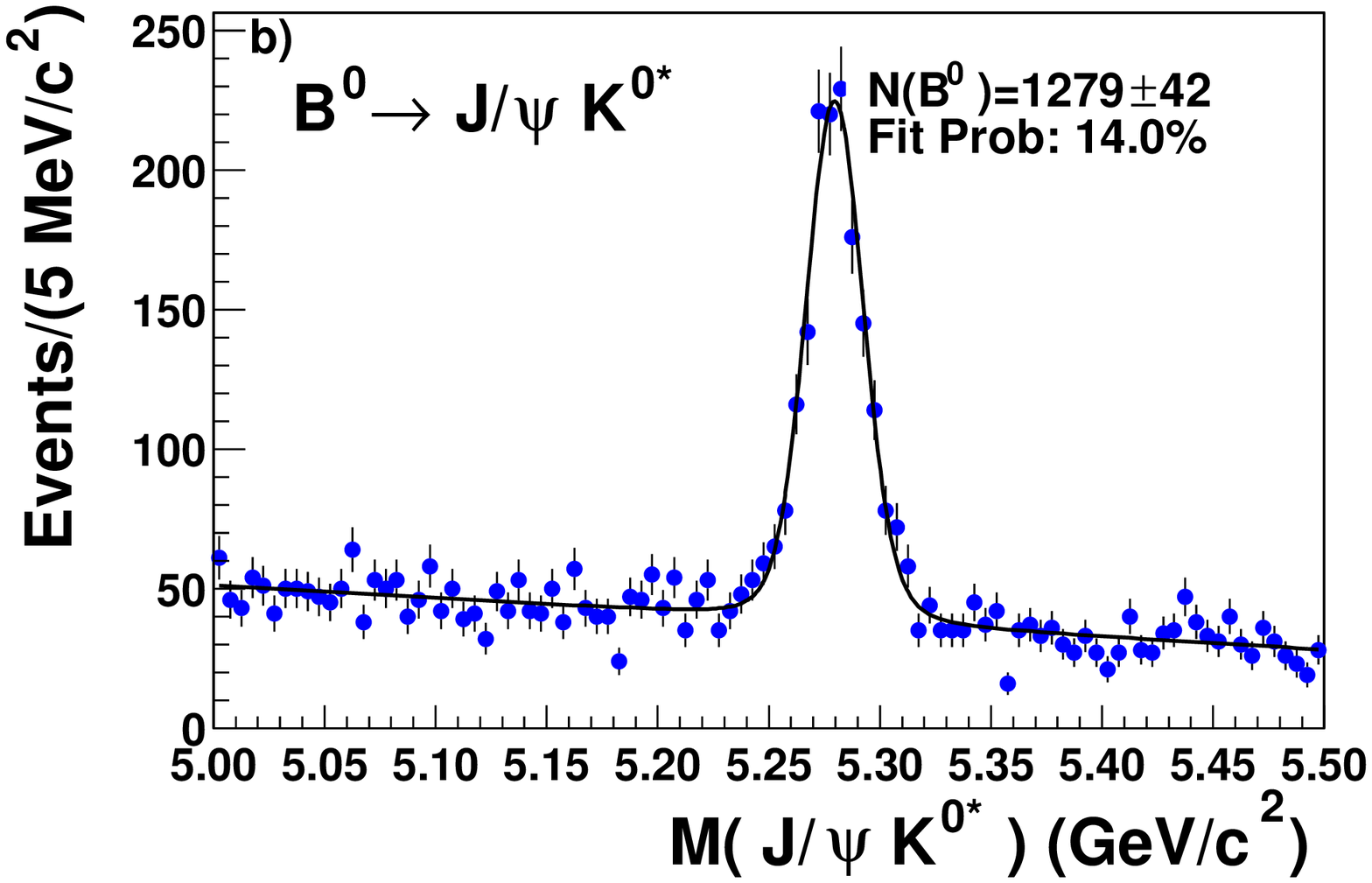}
 \end{minipage}\\
 \begin{minipage}[!h]{0.4\textwidth}
   \includegraphics[width=\textwidth]{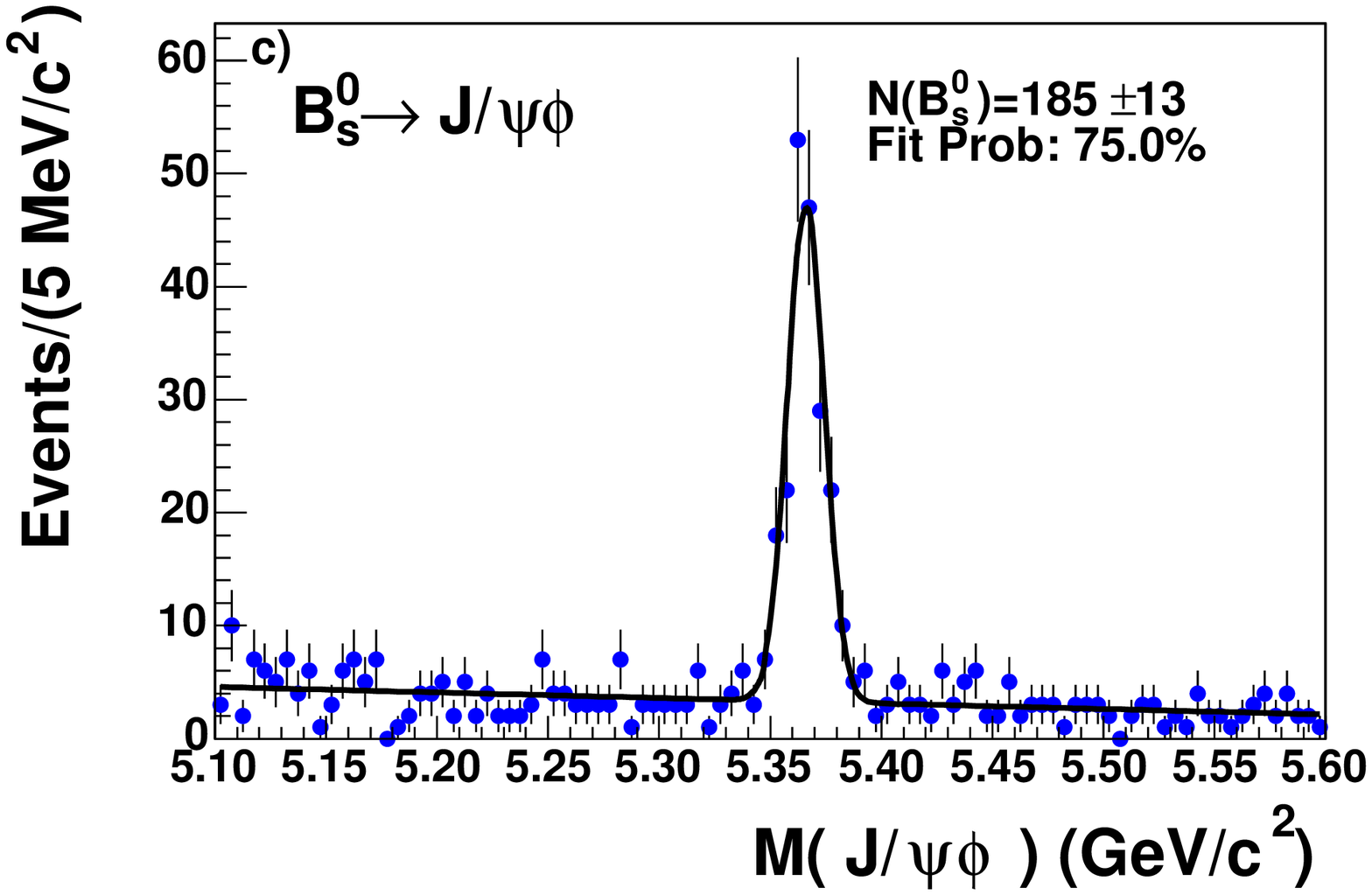}
 \end{minipage}
 \begin{minipage}[!h]{0.4\textwidth}
   \includegraphics[width=\textwidth]{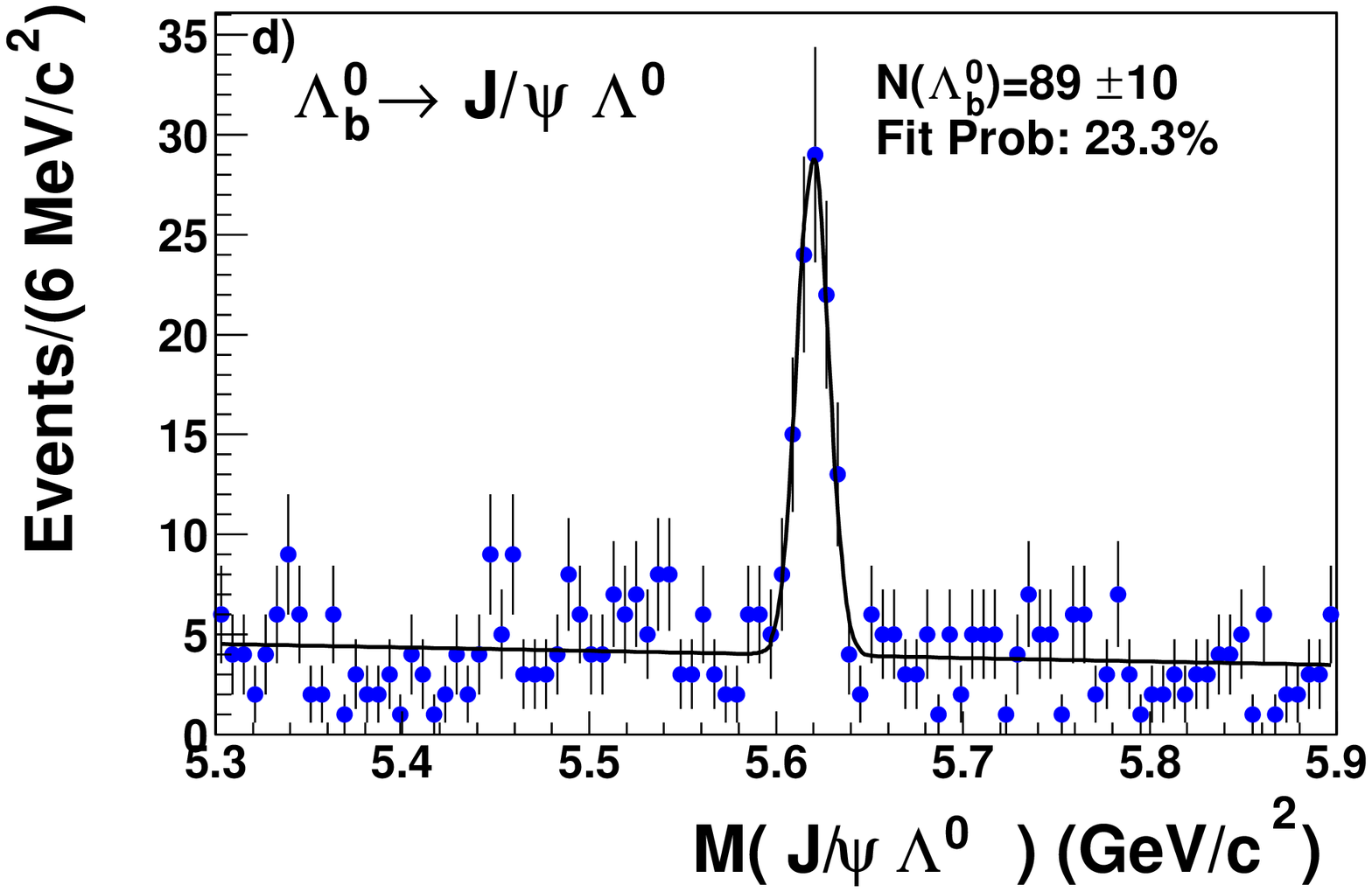}
 \end{minipage}
 \end{minipage}
 \end{center}
 \caption{The invariant mass distribution for $ J/\psi\;K^+$, $
 J/\psi\;K^{0*}$, $ J/\psi\;\phi$ and $ J/\psi\;\Lambda^0$ candidates.  
 The results of the log-likelihood fits are superimposed. The fit
 probability obtained from a $\chi^2$ test is shown.}
 \label{fig:Bu-Fit}
 \label{fig:BdFit}
 \label{fig:BsFit}
 \label{fig:LBFit}
 \end{figure*}

After reconstruction of candidates the mass is extracted using an
unbinned log-likelihood fit, with the signal distribution modeled 
as a Gaussian. The shape of the background is investigated using 
an inclusive Monte Carlo sample of $b$ hadron decays. 
A detailed detector simulation based on {\sc geant} is used.
The $B^+\to J/\psi\;K^+$ sample contains
significant contributions of partially reconstructed $B^0\to J/\psi
K^{*0}\to\mu^+\mu^- K^+\pi^-$ and misreconstructed $B^+\to J/\psi\;\pi^+$ 
decays, which are modeled in the background probability distribution function.
Partially reconstructed $B^0\to J/\psi K^{*0}\to\mu^+\mu^- K^+\pi^-$
decays populate the left shoulder in Figure~\ref{fig:Bu-Fit}a). 
Events from $B^+\to J/\psi\;\pi^+$ decays appear on the right side of the
signal peak. The misreconstruction of $K^{*0}\to K^+\pi^-$ due
to swapped track assignment of $K$ and $\pi$ is taken into account 
for $B^0\to J/\psi\;K^{*0}$ decays. 
No significant contributions are found for the other decay modes.
Comparisons between data and fits are shown in Figure~\ref{fig:Bu-Fit}.

The systematic uncertainties are summarized in 
Tables~\ref{tab:SystM} and~\ref{tab:SystLM}.  
The largest systematic uncertainties originate from the
momentum scale and tracker alignment.  
Deviations from the well-measured world averages in the $\psi'\to\mu^+\mu^-$,
$\psi'\to\mu^+\mu^-\pi^+\pi^-$ and $\Upsilon\to\mu^+\mu^-$ high
statistics samples are used to determine the uncertainty of the momentum scale.
The observed deviations, scaled by the Q-value of the respective decay,
provide an estimate of the systematic uncertainty.
The second uncertainty of importance originates from the relative 
alignment of SVX and COT.  It is evaluated by comparing 
mass measurements using the combined tracker information to
those using the COT information alone.
False curvature arises from tracker misalignments and results in a curvature
offset. The net observed effect is an increase in momentum for negatively
charged tracks and a decrease for positively charged tracks. 
False curvature effects cancel, to first order, in charge symmetric samples. 
We derive a parametric correction to remove the charge dependence
and use the mass shift due to this correction as a measure of the
systematic error. 
Uncertainties due to the vertex fit are evaluated using different
mass and pointing constraints in the fit.  
A correlation between fluctuations in measured curvature and
reconstructed vertex position is the source of the resolution 
bias uncertainty.  
Background systematics are determined by varying the background description.
In the $B^+$ case, the  mass shift due to inclusion and exclusion of the
misreconstructed $B^+\rightarrow J/\psi\pi^+$ is assigned as
systematic uncertainty. 
In the $B^0$ case, the systematic uncertainty is derived by varying the 
amount of the reflection contribution within $1 \sigma$ of expectation.

We obtain the following results: 
\[
\begin{array}{lcrcrcr}
m(B^+)           &=&5279.10&\pm&0.41\,_{(stat)}\;&\pm& 0.36\,_{(sys)}~\MeVcc,\\
m(B^0)\,         &=&5279.63&\pm&0.53\,_{(stat)}\;&\pm& 0.33\,_{(sys)}~\MeVcc,\\
m(B^0_s)\,       &=&5366.01&\pm&0.73\,_{(stat)}\;&\pm& 0.33\,_{(sys)}~\MeVcc,\\
m(\Lambda^0_b)\, &=&5619.7 &\pm&1.2 \,_{(stat)}\;&\pm& 1.2 \,_{(sys)}~\MeVcc.\\
\end{array}
\]
These results are in agreement with the current world averages:
$m(B^+)=5279.0\pm0.5~\MeVcc$, $m(B^0)=5279.4\pm0.5~\MeVcc$ and
$m(B^0_s)=5369.6\pm2.4~\MeVcc$~\cite{NEWPDG}.
Our new $\Lambda^0_b$ mass measurement agrees with two of the 
three previous measurements and is in excellent agreement 
with CDF's Run I measurement~\cite{LBRUNI,LBALEPH,LBDELPHI}.
The achieved precision is better than the current world average 
of $m(\Lambda^0_b) = 5624 \pm 9~\MeVcc$~\cite{NEWPDG}. 

\begin{table}[!h]
\begin{center}
  \caption{Summary of systematic uncertainties for the $B$ meson mass
  measurements in ${\rm MeV/}c^2$. }
  \label{tab:SystM}
\begin{ruledtabular}
\begin{tabular}{lllll}
\multicolumn{2}{l}{Source}  & $B^0\!\to\! J/\psi K^{*0}$
&$B^\pm\!\to\! J/\psi K^\pm$&  $B^0_s\!\to\! J/\psi\phi$\\ 
\multicolumn{2}{l}{Tracking }    & & &\\
~~& Momentum scale  &\hspace{5mm} 0.20        &\hspace{5mm} 0.22&\hspace{5mm} 0.20\\
  & Alignment       &\hspace{5mm} 0.18        &\hspace{5mm} 0.18\footnotemark[1]&\hspace{5mm} 0.18\footnotemark[1]\\
  & False Curvature &\hspace{5mm} 0.02\footnotemark[2]&\hspace{5mm} 0.02&\hspace{5mm} 0.02\footnotemark[2]\\ 
  & Vertex Fitting  &\hspace{5mm} 0.10        &\hspace{5mm} 0.10\footnotemark[1]&\hspace{5mm} 0.10\footnotemark[1]\\ 
  & Resolution bias &\hspace{5mm} 0.13        &\hspace{5mm} 0.13&\hspace{5mm} 0.13\\ \hline
\multicolumn{2}{l}{Bkg Systematics} &  & &\\
  & $K$-$\pi$ swap in $K^{*0}$&\hspace{5mm} 0.06&\hspace{5mm} ---  &\hspace{5mm} ---\\ 
  & $ J/\psi\pi$ contamin.&\hspace{5mm} --- &\hspace{5mm} 0.13  &\hspace{5mm} ---\\ 
\hline
\multicolumn{2}{l}{Total Uncertainty} 
                           &\hspace{5mm} 0.33  &\hspace{5mm} 0.36 &\hspace{5mm} 0.33 \\  
\end{tabular}
\end{ruledtabular}
\footnotetext[1]{from $B^0$}
\footnotetext[2]{from $B^\pm$}
\end{center}
\end{table}

\begin{table}[!h]
\begin{center}
  \caption{Summary of systematic uncertainties for the $\Lambda^0_b$ mass
    measurement in ${\rm MeV/}c^2$. The high statistics $B^0$
    values have been used for the $\Lambda^0_b$ systematics.}
  \label{tab:SystLM}
\begin{ruledtabular}
\begin{tabular}{llll}
\multicolumn{2}{l}{Source} &$B^0\to J/\psi K^0_S$ &  $\Lambda^0_b\to J/\psi\Lambda^0$ \\ 
\multicolumn{2}{l}{Tracking } & & \\
~~& Momentum scale    &\hspace{7mm} 0.2      &\hspace{7mm} 0.2\\
  & Alignment         &\hspace{7mm} 1.0      &\hspace{7mm} 1.0\footnotemark[1]\\ \hline
  & Vertex Fitting    &\hspace{7mm} 0.7      &\hspace{7mm} 0.7\footnotemark[1]\\ \hline
\multicolumn{2}{l}{Total uncertainty} 
                      &\hspace{7mm} 1.2      &\hspace{7mm} 1.2\\
\end{tabular}
\end{ruledtabular}
\footnotetext[1]{from $B^0\to J/\psi K^0_S$}
\end{center}
\end{table}

\begin{table}[tb]
\begin{center}
  \caption{Summary of systematic uncertainties for the $b$ hadron mass
  differences in ${\rm MeV/}c^2$.}
  \label{tab:MassDiffUncert}
\begin{ruledtabular}
\begin{tabular}{lclccc}
\multicolumn{3}{l}{mass difference      }&mom. scale&fit model& total uncert.\\
$m(B^\pm)\!\!\!\!$       &-& $\!\!\!\!m(B^0)$    &$0.00$&$0.14$&$0.14$\\ 
$m(B^0_s)\!\!\!\!$       &-& $\!\!\!\!m(B^0)$    &$0.01$&$0.06$&$0.06$\\
$m(B^0_s)\!\!\!\!$       &-& $\!\!\!\!m(B^\pm)$  &$0.01$&$0.13$&$0.13$\\
$m(\Lambda^0_b)\!\!\!\!$ &-& $\!\!\!\!m(B^0)$    &$0.05$&-     &$0.05$\\
\end{tabular}
\end{ruledtabular}
\end{center}
\end{table}

For the mass differences, most systematic uncertainties cancel.  
The momentum scale uncertainty is scaled down to the size of 
the mass difference. The remaining systematic uncertainty originates 
from differences in the fit models. 
We minimize systematic effects in $m(\Lambda^0_b)-m(B^0)$ by using the 
$B^0\to J/\psi K^0_S$ decay mode, which is topologically similar 
to $\Lambda^0_b \to J/\psi \Lambda^0$, and where we measure 
$m(B^0) = 5280.46 \pm 0.63\,_{(stat)}$, in agreement with the 
more precise mass determination from the $B^0 \to J/\psi K^{*0}$  decay mode.
The uncertainties are summarized in Table~\ref{tab:MassDiffUncert}.
We obtained the following results for the 
mass differences:
\[
\begin{array}{lclcrcrcr}
m(B^\pm)&-&m(B^0)      &=& -0.53    &\pm& 0.67\, &\pm& 0.14\, ~\MeVcc,\\
m(B^0_s)&-&m(B^0)      &=& 86.38    &\pm& 0.90\, &\pm& 0.06\, ~\MeVcc,\\
m(\Lambda^0_b)&-&m(B^0)&=& 339.2\;\;&\pm& 1.4 \, &\pm& 0.1\, ~\MeVcc.
\end{array}
\]

These are the most precise measurements of $m(B^0_s)-m(B^0)$ and $m(\Lambda^0_b)-m(B^0)$ to date.

We thank the Fermilab staff and the technical staffs of the
participating institutions for their vital contributions. This work
was supported by the U.S. Department of Energy and National Science
Foundation; the Italian Istituto Nazionale di Fisica Nucleare; the
Ministry of Education, Culture, Sports, Science and Technology of
Japan; the Natural Sciences and Engineering Research Council of
Canada; the National Science Council of the Republic of China; the
Swiss National Science Foundation; the A.P. Sloan Foundation; the
Bundesministerium f\"ur Bildung und Forschung, Germany; the Korean
Science and Engineering Foundation and the Korean Research Foundation;
the Particle Physics and Astronomy Research Council and the Royal
Society, UK; the Russian Foundation for Basic Research; the Comision
Interministerial de Ciencia y Tecnologia, Spain; in part by the
European Community's Human Potential Programme under contract
HPRN-CT-2002-00292; and the Academy of Finland. 
\bibliography{prl}

\end{document}

%% file: author_list_cdf_feb_2005.tex

\affiliation{Institute of Physics, Academia Sinica, Taipei, Taiwan 11529, Republic of China }
\affiliation{Argonne National Laboratory, Argonne, Illinois 60439 }
\affiliation{Institut de Fisica d'Altes Energies, Universitat Autonoma de Barcelona, E-08193, Bellaterra (Barcelona), Spain }
\affiliation{Istituto Nazionale di Fisica Nucleare, University of Bologna, I-40127 Bologna, Italy }
\affiliation{Brandeis University, Waltham, Massachusetts 02254 }
\affiliation{University of California, Davis, Davis, California 95616 }
\affiliation{University of California, Los Angeles, Los Angeles, California 90024 }
\affiliation{University of California, San Diego, La Jolla, California 92093 }
\affiliation{University of California, Santa Barbara, Santa Barbara, California 93106 }
\affiliation{Instituto de Fisica de Cantabria, CSIC-University of Cantabria, 39005 Santander, Spain }
\affiliation{Carnegie Mellon University, Pittsburgh, PA 15213 }
\affiliation{Enrico Fermi Institute, University of Chicago, Chicago, Illinois 60637 }
\affiliation{Joint Institute for Nuclear Research, RU-141980 Dubna, Russia }
\affiliation{Duke University, Durham, North Carolina 27708 }
\affiliation{Fermi National Accelerator Laboratory, Batavia, Illinois 60510 }
\affiliation{University of Florida, Gainesville, Florida 32611 }
\affiliation{Laboratori Nazionali di Frascati, Istituto Nazionale di Fisica Nucleare, I-00044 Frascati, Italy }
\affiliation{University of Geneva, CH-1211 Geneva 4, Switzerland }
\affiliation{Glasgow University, Glasgow G12 8QQ, United Kingdom }
\affiliation{Harvard University, Cambridge, Massachusetts 02138 }
\affiliation{Division of High Energy Physics, Department of Physics, University of Helsinki and Helsinki Institute of Physics, FIN-00014, Helsinki, Finland }
\affiliation{Hiroshima University, Higashi-Hiroshima 724, Japan }
\affiliation{University of Illinois, Urbana, Illinois 61801 }
\affiliation{The Johns Hopkins University, Baltimore, Maryland 21218 }
\affiliation{Institut f\"ur Experimentelle Kernphysik, Universit\"at Karlsruhe, 76128 Karlsruhe, Germany }
\affiliation{High Energy Accelerator Research Organization (KEK), Tsukuba, Ibaraki 305, Japan }
\affiliation{Center for High Energy Physics: Kyungpook National University, Taegu 702-701; Seoul National University, Seoul 151-742; and SungKyunKwan University, Suwon 440-746; Korea }
\affiliation{Ernest Orlando Lawrence Berkeley National Laboratory, Berkeley, California 94720 }
\affiliation{University of Liverpool, Liverpool L69 7ZE, United Kingdom }
\affiliation{University College London, London WC1E 6BT, United Kingdom }
\affiliation{Massachusetts Institute of Technology, Cambridge, Massachusetts 02139 }
\affiliation{Institute of Particle Physics: McGill University, Montr\'eal, Canada H3A~2T8; and University of Toronto, Toronto, Canada M5S~1A7 }
\affiliation{University of Michigan, Ann Arbor, Michigan 48109 }
\affiliation{Michigan State University, East Lansing, Michigan 48824 }
\affiliation{Institution for Theoretical and Experimental Physics, ITEP, Moscow 117259, Russia }
\affiliation{University of New Mexico, Albuquerque, New Mexico 87131 }
\affiliation{Northwestern University, Evanston, Illinois 60208 }
\affiliation{The Ohio State University, Columbus, Ohio 43210 }
\affiliation{Okayama University, Okayama 700-8530, Japan }
\affiliation{Osaka City University, Osaka 588, Japan }
\affiliation{University of Oxford, Oxford OX1 3RH, United Kingdom }
\affiliation{University of Padova, Istituto Nazionale di Fisica Nucleare, Sezione di Padova-Trento, I-35131 Padova, Italy }
\affiliation{University of Pennsylvania, Philadelphia, Pennsylvania 19104 }
\affiliation{Istituto Nazionale di Fisica Nucleare Pisa, Universities of Pisa, Siena and Scuola Normale Superiore, I-56127 Pisa, Italy }
\affiliation{University of Pittsburgh, Pittsburgh, Pennsylvania 15260 }
\affiliation{Purdue University, West Lafayette, Indiana 47907 }
\affiliation{University of Rochester, Rochester, New York 14627 }
\affiliation{The Rockefeller University, New York, New York 10021 }
\affiliation{Istituto Nazionale di Fisica Nucleare, Sezione di Roma 1, University di Roma ``La Sapienza," I-00185 Roma, Italy }
\affiliation{Rutgers University, Piscataway, New Jersey 08855 }
\affiliation{Texas A\&M University, College Station, Texas 77843 }
\affiliation{Texas Tech University, Lubbock, Texas 79409 }
\affiliation{Istituto Nazionale di Fisica Nucleare, University of Trieste/\ Udine, Italy }
\affiliation{University of Tsukuba, Tsukuba, Ibaraki 305, Japan }
\affiliation{Tufts University, Medford, Massachusetts 02155 }
\affiliation{Waseda University, Tokyo 169, Japan }
\affiliation{Wayne State University, Detroit, Michigan 48201 }
\affiliation{University of Wisconsin, Madison, Wisconsin 53706 }
\affiliation{Yale University, New Haven, Connecticut 06520 }


\author{D.~Acosta}
\affiliation{University of Florida, Gainesville, Florida 32611 }

\author{J.~Adelman}
\affiliation{Enrico Fermi Institute, University of Chicago, Chicago, Illinois 60637 }

\author{T.~Affolder}
\affiliation{University of California, Santa Barbara, Santa Barbara, California 93106 }

\author{T.~Akimoto}
\affiliation{University of Tsukuba, Tsukuba, Ibaraki 305, Japan }

\author{M.G.~Albrow}
\affiliation{Fermi National Accelerator Laboratory, Batavia, Illinois 60510 }

\author{D.~Ambrose}
\affiliation{Fermi National Accelerator Laboratory, Batavia, Illinois 60510 }

\author{S.~Amerio}
\affiliation{University of Padova, Istituto Nazionale di Fisica Nucleare, Sezione di Padova-Trento, I-35131 Padova, Italy }

\author{D.~Amidei}
\affiliation{University of Michigan, Ann Arbor, Michigan 48109 }

\author{A.~Anastassov}
\affiliation{Rutgers University, Piscataway, New Jersey 08855 }

\author{K.~Anikeev}
\affiliation{Fermi National Accelerator Laboratory, Batavia, Illinois 60510 }

% ...
